\documentstyle[11pt,newpasp,twoside]{article}
\def\edcomment#1{\iffalse\marginpar{\raggedright\sl#1\/}\else\relax\fi}
\reversemarginpar

\begin{document}
\title{Blue compact galaxies and the primordial $^4$Helium abundance  }
 \author{Trinh Xuan Thuan}
\affil{Astronomy Department, University of Virginia, 
P.O. Box 3818, University Station, Charlottesville, VA 22903, USA}
\author{Yuri I. Izotov}
\affil{Main Astronomical Observatory, Ukrainian National Academy of Sciences, 
Goloseevo, Kiev 03680, Ukraine}

\begin{abstract}

Blue compact galaxies (BCG) are ideal objects in which to derive the 
primordial 
$^4$He abundance because they are chemically young and have not had a 
significant stellar He contribution. We discuss a self-consistent method 
which makes use of all the brightest He I emission lines in the optical 
range and solves consistently for the electron density of the He II zone.
We pay particular attention to electron collision and radiative transfer 
as well as underlying stellar absorption effects which may make the 
He I emission lines deviate from their recombination values. 
 Using a large homogeneous sample of 45 low-metallicity H II regions 
in BCGs, and  extrapolating the Y-O/H and Y-N/H linear regressions to 
O/H = N/H = 0, we obtain Y$_p$ = 0.2443$\pm$0.0015, in excellent agreement 
with the weighted mean value Y$_p$ = 0.2452$\pm$0.0015
obtained from the detailed analysis of the two most 
metal-deficient BCGs known, I Zw 18 and SBS 0335--052. The derived slope 
dY/dZ = 2.4$\pm$1.0 is in agreement with the value derived for the Milky Way 
and with simple chemical evolution models with homogeneous outflows.
Adopting $Y_p$ = 0.2452$\pm$0.0015
leads to a baryon-to-photon ratio of 
(4.7$^{+1.0}_{-0.8}$)$\times$10$^{-10}$ and to a
baryon mass fraction in the Universe $\Omega_b$$h^2_{50}$ = 
0.068$^{+0.015}_{-0.012}$,
consistent with the value derived from the primordial D abundance of 
 Burles \& Tytler (1998).

\end{abstract}

\section{Introduction}

Blue compact galaxies (BCG) are low-luminosity (M$_B$ $\ga$ -18) 
systems which are undergoing an intense burst of star formation in a very 
compact region (less than 1 kpc) which dominates the light of the galaxy 
(Figure 1) and which shows blue colors and a HII region-like emission-line 
optical spectrum (Figure 2). BCGs are ideal laboratories in which to measure 
the primordial $^4$Helium abundance because of several reasons:

1) With an oxygen abundance O/H ranging between 1/50 and 1/3 that of the 
Sun, BCGs are among the most metal-deficient gas-rich galaxies known.
Their gas has not been processed through many generations of stars, and thus   
best approximates the pristine primordial gas. Izotov \& Thuan (1999) have 
argued that BCGs with O/H less than $\sim$ 1/20 that of the Sun may be genuine 
young galaxies. Their 
argument is based on the observed constancy and very small scatter of the C/O 
and N/O ratios in extremely metal-deficient BCGs with 12 + log O/H $\la$ 
7.6, which they 
interpret as implying that the C and N in these galaxies have been made in 
the same massive stars (M $\ga$ 9 M$_\odot$) which manufactured O,
 and that intermediate-mass stars 
(3 M$_\odot$ $\la$ M $\la$ 9 M$_\odot$) have not had time to release their 
nucleosynthetic products. Since the main-sequence lifetime of a 9 M$_\odot$ 
star is $\sim$ 40 Myr, Izotov \& Thuan (1999) suggest that very 
metal-deficient BCGs are younger than $\sim$ 100 Myr. Thus the primordial 
Helium mass fraction Y$_p$ can be derived accurately in very 
metal-deficient BCGs with only a small correction for Helium made in stars.

%
%  Slit orientations in I Zw 18 and SBS 0335-052 (fig.1)
%
\begin{figure}
\vspace{3.5cm}
%\figurenum{1}
%\epsscale{1.6}
%\plotfiddle{thuan_fig1.ps}{3.5cm}{270.}{45.}{45.}{-180.}{170.}
%\plotone{fig1.ps}
\vspace{2.cm}
%\vspace{2.5cm}
\caption{Slit orientations superposed on {\sl HST}
archival $V$ images of I Zw 18 and SBS 0335--052. The slit 
orientation of I Zw 18 is chosen in such a way as to get 
spectra of the SE and NW components as well as of the C component to the 
NW of the main body of the galaxy. 
The spatial scale is 1\arcsec\ = 49 pc in the case of I Zw 18 and is
1\arcsec\ = 257 pc in the case of SBS 0335--052.}
\end{figure}

2) Because of the relative insensitivity of $^4$He production to the 
baryonic density of matter,  Y$_p$ needs to be 
determined to a precision better than 5\% to provide useful cosmological 
constraints. This precision can in principle be achieved by using BCGs because
their optical spectra show several He I recombination emission lines and  
very high signal-to-noise ratio emission-line spectra with moderate spectral 
resolution of BCGs can be obtained at large telescospes (4 m class or larger) 
coupled with efficient and linear CCD detectors with a relatively modest 
investment of telescope time.
 The theory of nebular emission is well understood and the theoretical He I 
recombination coefficients calculated by 
Brocklehurst (1972) and Smits (1996) are well known enough to allow to 
convert He emission-line strengths into abundances with the desired accuracy.

\section{The primordial He abundance from extrapolation of the Y-O/H and 
Y-N/H linear regressions}

\subsection{A new large sample of Blue Compact Galaxies}

%
%   Spectrum of I Zw 18 (fig.2)
%
\begin{figure}
%\figurenum{2}
%\epsscale{2.0}
\plotfiddle{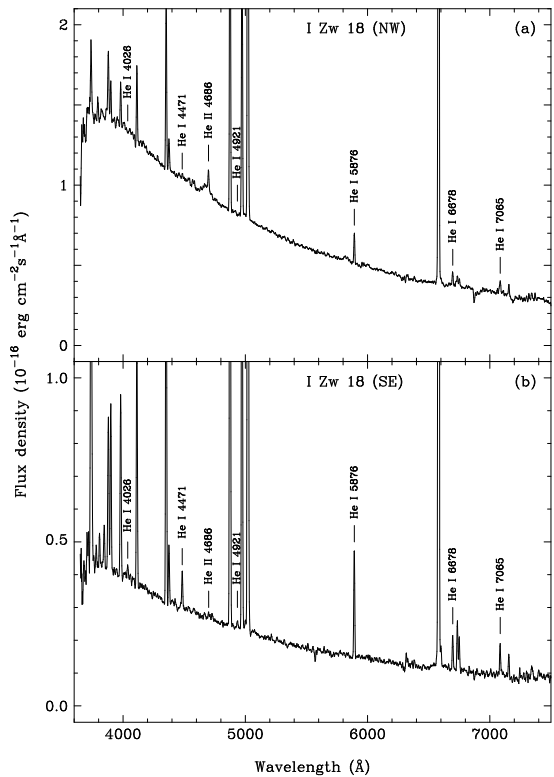}{5.5cm}{0.}{40.}{40.}{-120.}{-150.}
%\plotone{fig2.ps}
\vspace{4.5cm}
\caption{The 0\farcs6$\times$1\farcs5 aperture MMT spectra
of the brightest parts of the NW and the SE components of I Zw 18. The spectrum
of the SE component is extracted at the angular distance of 5\farcs4 from
the NW component. The positions of He I lines
are marked. Note that all marked He I lines in the spectrum of the SE
component are in emission while the two He I $\lambda$4026 and $\lambda$4921
lines are in absorption and the He I $\lambda$4471 emission line is 
barely detected in the spectrum of the NW component.}
\end{figure}

Peimbert \& Torres-Peimbert (1974, 1976) first noted the correlation between 
He and O abundances in a small sample of dwarf magellanic irregulars and 
BCGs, and they proposed to determine Y$_p$  by 
linear extrapolation of the correlation to O/H = 0. Later, Pagel, Terlevich 
\& Melnick (1986) proposed to use also the Y-N/H correlation for the 
determination of  
Y$_p$, to take into account the temporary local excess of helium and nitrogen 
due to pollution by winds from massive stars. Many attempts at determining 
Y$_p$ have been made, using the Y versus O/H and Y versus N/H correlations 
on various samples of dwarf irregulars and BCGs (e.g.
Pagel et al. 1992, Izotov, Thuan \& Lipovetsky (1994, 1997, hereafter ITL94 
and ITL97; Olive, Steigman \& SKillman 1997, hereafter OSS97; Izotov \& Thuan 
1998ab, hereafter IT98ab).

Before our work, the largest, most accurate and consistent published data 
set was by Pagel et al. (1992). Their observations were reduced in a uniform 
manner and they paid careful attention to such points as 
the correction for the unseen neutral helium and electron 
collisional effects which may make some He I lines deviate from their 
recombination values. Pagel et al. (1992) 
obtained Y$_p$ = 0.228$\pm$0.005, below the limit set by the standard 
hot big bang model of nucleosynthesis (SBBN) and consistent with it only at 
the 2$\sigma$ level. This prompted us to consider obtaining 
another measurement of Y$_p$ from an independent data set with as high 
or better precision to test SBBN.

Starting in 1993, we embarked on a large-scale program to obtain high 
signal-to-noise ratio spectra for a relatively large sample of BCGs 
 assembled from several objective-prism surveys: the First Byurakan or 
Markarian survey (Markarian et al. 1989), the Second Byurakan 
Survey (SBS, Izotov et al. 1993) and the University of Michigan survey 
(Salzer, MacAlpine \& Boroson 1989).
The SBS sample was particularly interesting because it contained about 
a dozen BCGs with O/H less than 1/15 of (O/H)$_\odot$, more than doubling the 
number of such known low-metallicity BCGs. The total sample consists of
45 H II regions in 42 BCGs. The data have been published 
in a series of papers in the Astrophysical Journal: ITL94, ITL97 and IT98ab. 

\subsection{Methodology}

There are a number of features which distinguish 
our work from previous efforts in determining the primordial He abundance.
Our methodology is described in detail in ITL94, ITL97, and IT98ab.

1) We have observed all the galaxies in our sample with the same 
telescopes (the Kitt Peak 4 m and 2.1 m telescopes) and instrumental set up,
and the data were all reduced in a homogeneous way. 
This differs from OSS97, for 
example, which used a more heterogeneous sample of BCGs observed by different
 observers on different telescopes, with some of the data obtained many years 
ago with nonlinear detectors. A uniform sample is essential to minimize as 
much as possible the artificial scatter introduced by assembling different 
data sets reduced in different ways.

2) To derive the He mass fraction, previous authors use mainly one 
He emission line, He I 6678. Correction to this line's intensity 
is usually made only for one effect, electron collisional enhancement. This  
correction is usually carried out adopting 
the electron density derived from the [S II] 6717/6731 emission-line ratio.
The approach just described has several shortcomings. 
The metastability of the 2$^3$S state of He I can also lead to possible 
radiative 
transfer effects (also called fluorescence effects) in the triplet lines 
which may be enhanced at the expense of the He I 3889 line (Robbins 1968).
When a single He I emission 
line is used, one cannot distinguish between between electron collisional 
and radiative transfer effects. Thus, fluorescent enhancement is neglected, 
while it may be important. Furthermore,
 at the low electron number densities N$_e$ 
which often characterize the HII regions in BCGs, the 
determination of N$_e$ from [S II] emission lines is very uncertain. In the 
majority of cases, N$_e$ is arbitrarily set to 100 cm$^{-3}$. This 
assumption can lead to artificially low He abundance, as in the case of the
southeast component of I Zw 18  where the true  N$_e$ is only 
$\sim$ 10 cm$^{-3}$ (Izotov et al. 1999).
 More importantly, setting  N$_e$(He II) equal to  
N$_e$(S II) is not physically reasonable as the S$^+$ and He$^+$ regions are 
not expected to coincide, given the large difference in the S I and He I
 ionization potentials.

%
%   Spatial distribution of He I line equivalent widths (fig.3)
%
\begin{figure}
%\figurenum{6}
%\epsscale{2.0}
\plotfiddle{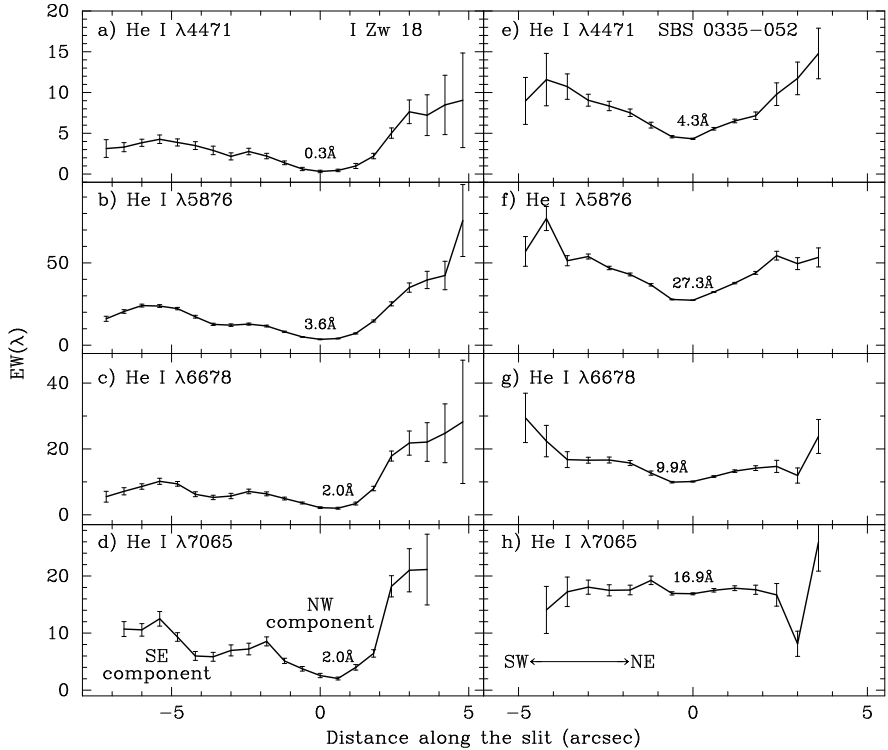}{7.5cm}{270.}{50.}{50.}{-190.}{230.}
%\plotone{fig3.ps}
\vspace{2.cm}
\caption{The spatial distributions of the He I nebular emission 
line equivalent widths in I Zw 18 (left panel) and in SBS 0335--052 (right 
panel). The error bars are 1$\sigma$ deviations. The value of the minimum 
equivalent width for each He I emission line is given.}
\end{figure}

To remedy these problems, we have proposed a self-consistent method 
in which we use all five brightest He I emission-lines in the 
optical range (the He I 3889, 4471, 5876, 6678 and 7065 lines) and solve 
simultaneously for N$_e$(He II) and the optical depth in the He I 3889 line 
so that the He I 3889/4471, 5876/4471, 6678/4471 and 7065/4471 line ratios 
have their recombination values, after correction for both collisional 
(Kingdon \& Ferland 1995) and fluorescent (Robbins 1968) enhancements. The 
He I 3889 and 7065 lines play an important 
role because they are particularly sensitive to both optical depth and 
electron number density.

\subsection{Underlying stellar absorption}

Effects other than collisional and fluorescent enhancements can also change 
He I line intensities. An important effect is the underlying stellar 
absorption in He I lines caused by hot OB stars which decreases the 
intensities of nebular He I lines. This effect is most important for the 
emission lines with the smallest equivalent widths.

%
%   Linear regression (fig.4)
%
\begin{figure}
%\figurenum{2}
%\epsscale{2.0}
\plotfiddle{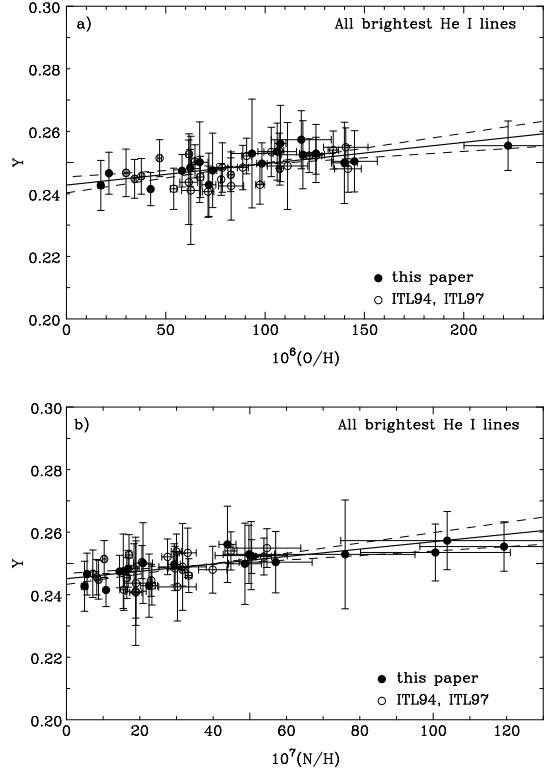}{5.5cm}{0.}{40.}{40.}{-120.}{-150.}
%\plotone{fig2.ps}
\vspace{4.5cm}
\caption{Linear regressions of (a) the helium mass fraction
$Y$ vs. oxygen abundance O/H and (b) the helium mass fraction $Y$ vs.
nitrogen abundance for our sample of
45 H II regions. The $Y$s are derived self-consistently 
by using the 5 brightest He I emission lines in the optical range.
Collisional and fluorescent enhancements, underlying He I stellar absorption and
Galactic Na I interstellar absorption are taken into account.
Open circles denote data from ITL94 and ITL97 and filled circles are data from
IT98ab. 1$\sigma$ alternatives are
shown by dashed lines.}
\end{figure}

The neglect of He I underlying stellar absorption can lead to a severe 
underestimate of the He mass fraction. One of the most spectacular examples 
is that of the BCG I Zw18. This object plays a key role in the determination 
of the primordial He abundance because, with an O/H only 1/50 that of the 
Sun, it is the most metal-deficient BCG known and has great influence on the
derived slopes and intercepts of the Y-O/H and Y-N/H linear regression lines.
Figure 2 shows very high signal-to-noise ratio spectrophotometric observations
of I Zw 18 obtained with the Multiple Mirror Telescope (MMT),  with the slit 
oriented so as to go through the two main centers of star formation in the 
BCG, the so-called NW and SE components (Figure 1, left). Comparison of 
the spectrum of the NW component (Figure 2a) with that of the SE component 
(Figure 2b) shows clearly that underlying stellar absorption is much more 
important in the NW than in the SE component: all marked He I lines in the 
spectrum of the SE component are in emission while the two He I 4026 and 4921 
lines are in absorption and the He I 4471 emission line is barely visible 
in the spectrum of the NW component. Prior to our work, most of the He work 
relied on measurements of Y in the NW component because of its larger 
brightness as compared to the SE component (Figure 1, left). This led to 
a systematic underestimate of Y in I Zw 18 . 
Izotov et al. (1999) derive the implausibly 
small Y(4471)= 0.169$\pm$0.023  and Y(5876)= 0.192$\pm$0.007 from the 
He I 4471 and 5876 lines respectively. In addition to underlying stellar 
absorption, the 5876 line intensity in I Zw 18 is reduced further 
by absorption from the Galactic interstellar 5890 and 5896 Na I lines.

%
%   Spatial distribution of EW (fig.5)
%
\begin{figure}
%\figurenum{6}
%\epsscale{2.0}
\plotfiddle{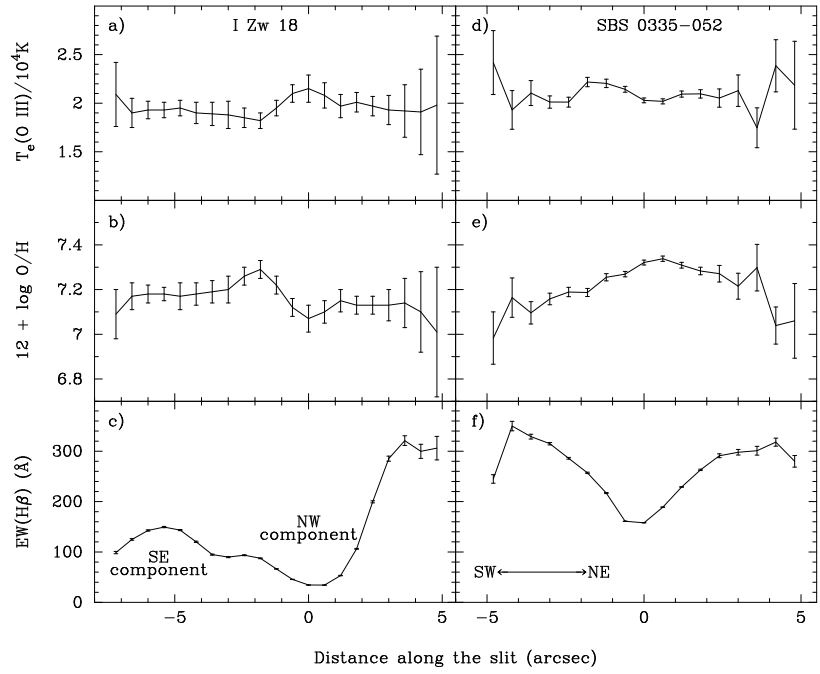}{7.5cm}{270.}{45.}{45.}{-180.}{230.}
%\plotone{fig3.ps}
\vspace{1.cm}
\caption{The spatial distributions of the electron temperature
$T_e$(O III), oxygen abundance 12 + log (O/H), and the equivalent width 
$EW$(H$\beta$) of the H$\beta$ emission line in I Zw 18 (left panel) and in
SBS 0335--052 (right panel). The error bars are 1$\sigma$ deviations.}
\end{figure}

 While the NW component cannot be used for He determination, the SE 
component is better suited as the influence of 
underlying stellar absorption on the He I emission line intensities is 
significantly smaller in this component. This can be seen in 
the left panel of Figure 3 which shows the spatial distributions of the 
He I nebular emission-line equivalent widths in I Zw 18: they are 
systematically larger in the SE component than in the NW one, implying 
less absorption. 

\subsection{Results}

Figure 4 shows the Y-O/H and Y-N/H linear regressions for the whole sample 
of 45 H II regions in BCGs. The sample includes most of the very 
metal-deficient BCGs known, including the two most extreme ones, the 
SE component of I Zw 18 and SBS 0335--052 with O/H about 1/43 that of the 
Sun. We obtain Y$_p$= 0.2443$\pm$0.0015 with dY/dZ = 2.4$\pm$1.0 (IT98b).
 Our Y$_p$ 
is considerably higher than those derived by other groups which range 
from 0.228$\pm$0.005 (Pagel et al. 1992) to 0.234$\pm$0.002 (OSS97). 
At the same time, our derived slope is significantly smaller than those of 
other authors, dY/dZ = 6.7$\pm$2.3 for Pagel et al. (1992) and 
dY/dZ = 6.9$\pm$1.5 for OSS97. This shallower slope is in good agreement 
with the value derived from stellar data for the Milky Way's disk  
 and with simple models of galactic evolution of BCGs with well-mixed 
homogeneous outflows.

%
%   Spatial distrib of Y in SBS 0335-052 (fig.6)
%
\begin{figure}
%\figurenum{2}
%\epsscale{2.0}
\plotfiddle{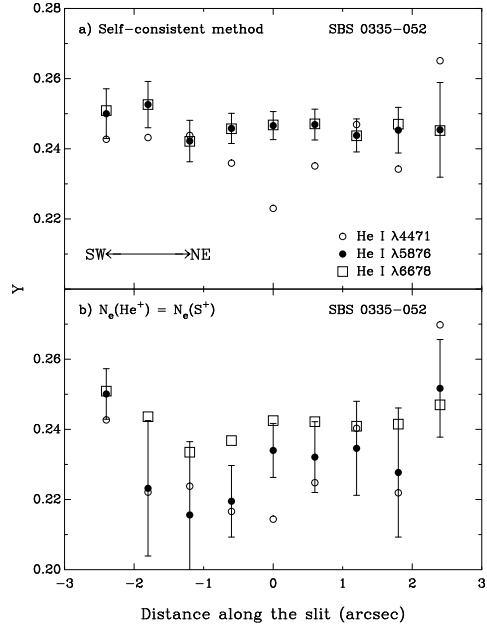}{4.cm}{0.}{35.}{35.}{-120.}{-120.}
%\plotone{fig2.ps}
\vspace{3.5cm}
\caption{The spatial distributions of the helium mass fractions
in SBS 0335--052 derived from the He I $\lambda$4471, $\lambda$5876
and $\lambda$6678 emission line intensities. The intensities of the He I
emission lines in Figure 6a are corrected for fluorescent and collisional
enhancement with an electron number density $N_e$(He II) and an optical depth
$\tau$($\lambda$3889) derived self-consistently from the observed 
He I $\lambda$3889, $\lambda$4471, $\lambda$5876, $\lambda$6678 and 
$\lambda$7065 emission line intensities. The intensities of the He I emission
lines in Figure 6b are corrected only for collisional enhancement with an
electron number density $N_e$(S II). The 1$\sigma$ error bars are shown only for 
the He mass fraction derived from the He I $\lambda$5876 emission line. They are 
larger in Figure 6b because of the large uncertainties in the determination of 
$N_e$(S II).}
\end{figure}

\section{He abundance in the two most metal-deficient blue compact 
galaxies known}

\subsection{I Zw 18 and SBS 0335--052}   

Instead of the statistical approach described above, we can also derive 
the primordial He abundance from accurate measurements of the He 
abundance in  a few objects selected to have very low O/H 
to minimize the amount of He manufactured in stars.

Izotov et al. (1999) 
have carried out such a study for the two most metal-deficient BCGs known.
I Zw 18 and SBS 0335--052 provide a study in contrast concerning the 
different physical mechanisms which may modify the 
He I emission-line intensities. While in I Zw 18, the electron number density 
is small (N$_e$ $\la$ 100 cm$^{-3}$) and collisional enhancement has a 
minor effect on the derived helium abundance, N$_e$ is much higher 
in SBS 0335--052 (N$_e$ 
$\sim$ 500 cm$^{-3}$ in the central part of the H II region). Additionally, 
the linear size of the H II region in SBS 0335--052 is $\sim$ 5 times 
larger than in I Zw 18, suggesting that it may be optically thick for 
some He I transitions. In fact, both collisional and fluorescent 
enhancements of He I emission lines play an important role in this 
galaxy. By contrast, underlying stellar He I absorption is much less 
important in SBS 0335--052 than in I Zw 18. Since the 
equivalent widths (EW) of the He I emission lines scale roughly as the H$\beta$
EWs, it is evident that this is the case from Figures 5c and 5f which 
show the spatial distribution of EW(H$\beta$) in both BCGs.
In SBS 0335--052, EW(H$\beta$) has a lowest value of 160 \AA\ and increases to 
$\sim$ 300 \AA\ in the outer parts, while in I Zw 18 , EW(H$\beta$) is only 
34 \AA\ in the center of the NW component. Given equal EWs for He I absorption 
lines in both BCGs, we may expect the effect of underlying stellar 
absorption to be 
$\sim$ 5 times smaller in SBS 0335--052 than in I Zw 18.

To disentangle the various effects which may make the He I emission-line 
intensities deviate from their recombination values, it is thus essential to 
use as many He I lines as possible in a self-consistent method as 
decribed above. An important and essential check that all corrections have 
been properly applied is the agreement between the He mass fraction Y derived 
independently from each He line. Figure 6 shows the Ys derived from the 
4471, 5876 and 6678 He I emission lines in  SBS 0335--052 at different spatial 
locations. It is clear that the self-consistent method (Figure 6a) gives 
much better agreement between the different lines. The Ys derived from 
the 4471 line are systematically below because only collisional and 
fluorescence effects have been taken into account and not underlying stellar 
absorption, and because the 4471 line is more subject to the latter effect.
By contrast, there is not very good agreement between the Ys from different 
lines when  N$_e$(He II) is set equal to N$_e$(S II) and only collisional 
enhancement is taken into account (Figure 6b). 

\subsection{Results}   
    
Izotov et al. (1999) derive Y = 0.243$\pm$0.007 for the SE component 
of I Zw 18 in very good agreement wth the value found by IT98a and Y = 
0.2463$\pm$0.0015 for SBS 0335--052, excluding the He I 4471 line. The 
weighted mean is then Y = 0.2462$\pm$0.0015. Using dY/dZ = 2.4 (IT98b), the 
stellar He contribution is 0.0010, giving a primordial value Y$_p$ = 
0.2452$\pm$0.0015, in excellent agreement with the value 0.2443$\pm$0.0015 
derived 
from extrapolation of the Y-O/H and Y-N/H regression lines for our 
large BCG sample. 
It is, however, higher than the value Y$p$ = 0.2345$\pm$0.0030 
derived by Peimbert \& Peimbert (2000) from Magellanic Clouds H II regions.
These authors suggest that two systematic effects may cause the 
disagreement: 
the presence of neutral hydrogen inside the helium Stromgren sphere and 
temperature fluctuations in our BCGs.

%
%   He vs eta (fig.7)
%
\begin{figure}
%\figurenum{2}
%\epsscale{2.0}
\plotfiddle{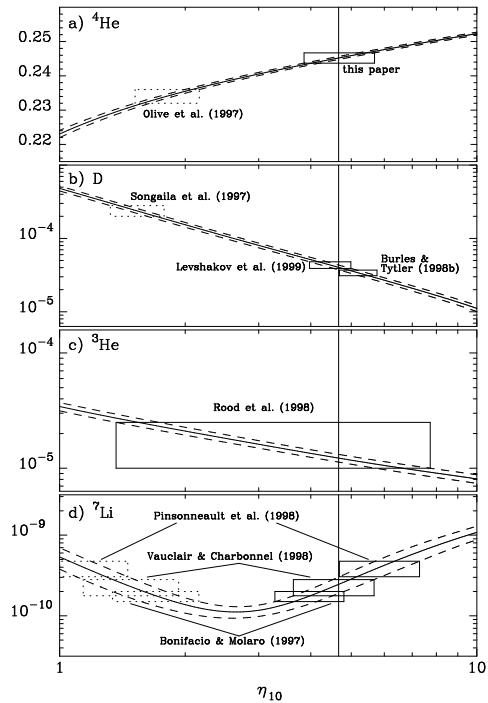}{5.cm}{0.}{35.}{35.}{-120.}{-120.}
%\plotone{fig2.ps}
\vspace{4.cm}
\caption{The abundance of (a) $^4$He, (b) D, (c) $^3$He 
and (d) $^7$Li
as a function of $\eta_{10}$ $\equiv$ 10$^{10}$ $\eta$, where $\eta$ is the 
baryon-to-photon 
number ratio, as given by the standard hot big bang nucleosynthesis model.
%The theoretical dependences (solid lines) with 1$\sigma$ deviations
%(dashed lines) are from Fiorentini et al. (1998). 
The abundances of D, $^3$He and $^7$Li are number ratios
relative to H. For $^4$He, the mass fraction $Y$ is shown. Our
value $Y_p$ = 0.2452$\pm$0.0015 gives $\eta$ = 
(4.7$^{+1.0}_{-0.8}$)$\times$10$^{-10}$
as shown by the solid vertical line. We show other data with 1$\sigma$ boxes.}
\end{figure}

There is no evidence that the first effect is important in our objects.
In our work, we have used the 'radiation softness parameter' of 
Vilchez \& Pagel (1988) to estimate the correction factor for neutral helium
and found the fraction of neutral helium to be insignificant ($\la$ 2\%)
in all our objects, i.e their HII and He II Stromgren 
spheres are coincident to a very 
good approximation. This conclusion is corroborated by the detailed modeling 
of I Zw 18 by Stasinska \& Schaerer (1999) who found the amount of 
neutral Helium to be negligible. On the other hand, they did find  
the observed T$_e$(OIII) to be $\sim$ 15\% higher than predicted by 
the photoionization model.
Future progress in the determination of the primordial $^4$He 
abundance using BCGs will rely on the discovery of more I Zw 18-like objects 
and on detailed modeling of very high signal-to-noise 
ratio and high-spectral resolution spectra of a few very metal-deficient 
BCGs (those with O/H less than 1/20 of solar) to look into such systematic 
effects as those mentioned above. 

\section{Cosmological implications}

Figure 7 shows the primordial abundances of $^4$He, D, $^3$He, and $^7$Li 
predicted by standard big bang nucleosynthesis theory as a function of the 
baryon-to photon number ratio $\eta$. The dashed lines are 1 $\sigma$ 
uncertainties in model calculations. The solid boxes show the 1 $\sigma$ 
predictions of $\eta$ as inferred from our derived primordial 
abundance of $^4$He, and the primordial abundances of D (Levshakov, Kegel 
\& Takahara 1999; Burles \& Tytler 1998), $^3$He (Rood et al. 1998) and 
$^7$Li (Bonifacio \& Molaro 1997; Vauclair \& Charbonnel 1998; Pinsonneault 
et al. 1999). All these determinations are consistent to within 1 $\sigma$, 
although the most stringent constraint is provided by D. For comparison, 
we have also plotted the Y$_p$ derived by OSS97 which is low, partly 
because underlying stellar absorption was not taken into account in I Zw 18.
Their Y$_p$ = 0.234 happens also to be the value obtained by Peimbert \& 
Peimbert (2000) in the Magellanic Clouds. This low value would 
have been consistent with the primordial D abundance obtained by Songaila et 
al (1997) which is one order of magnitude higher than the value obtained by 
Burles \& Tytler (1998), except that this high value is now believed to be 
erroneous. 

Our Y$_p$ = 0.2452$\pm$0.0015 value implies a baryon-to-photon number ratio 
$\eta$ = 4.7$^{+1.0}_{-0.8}$ $\times$10$^{-10}$. This 
translates to a baryon mass fraction $\Omega_b$$h^2_{50}$ = 
0.068$^{+0.015}_{-0.012}$ where $h_{50}$ is the Hubble constant in units of 50 
km s$^{-1}$Mpc$^{-1}$. For a Hubble constant equal to 65 km s$^{-1}$Mpc$^{-1}$,
$\Omega_b$ = 0.040$^{+0.009}_{-0.007}$. Our 
derived baryonic mass fraction is  consistent with the one 
obtained by analysis of the Ly$\alpha$ forest in a 
 cold dark matter cosmology. Depending on the intensity
of diffuse UV radiation, the inferred lower limit is 
$\Omega$$h^2_{50}$ = 0.05 -- 0.10 ( Weinberg et al. 1997; Bi \& Davidsen 1997; 
Rauch et al. 1997), while Zhang et al. (1998) have derived 0.03 
$\leq$$\Omega$$h^2_{50}$ $\leq$ 0.08.

Finally, for the most consistent set of primordial abundances 
-- D  from Levshakov et al. (1999),
 our above value for $^4$He,
and $^7$Li from Vauclair \& Charbonnel (1998) --
 we derive an equivalent number of light neutrino
species $N_\nu$ = 3.0$\pm$0.3 (2$\sigma$) (Izotov et al. 1999).

\acknowledgments

We thank the partial financial support of NSF grant AST-96-16863 and an 
IAU Travel grant. We acknowledge useful conversations with M. Peimbert and 
B. Pagel and thank the organizers for a stimulating meeting in a 
superb locale.

\end{document}